# Fourier Spectrum Pulse-Echo for Acoustic Characterization

Barnana Pal

Saha Institute of nuclear Physics,
Condensed Matter Physics Division,
1/AF, Bidhannagar, Kolkata-700064, India.
e-mail: barnana.pal@saha.ac.in

Abstract:

Ultrasonic wave attenuation($\alpha$) measurement by pulse-echo method exhibits pronounced dependence on experimental conditions. It is shown to be an inherent characteristic of the method itself. Estimation of $\alpha$ from the component wave amplitudes in the frequency scale gives more accurate and consistent value. This technique, viz., the Fourier Spectrum Pulse-Echo (FSPE) is demonstrated to determine the ultrasonic velocity($v$) and attenuation constant($\alpha$) in ultrapure de-ionized water at room temperature ($25^0$C) at 1MHz and 2 MHz wave frequency.

Keywords: Fourier transform, pulse echo method, ultrasonic propagation constants.

## 1. Introduction:

Fourier transform spectroscopic techniques in experimental research have been growing at a faster rate during past five decades only because of the fact that higher precision and better accuracy in measurement can be achieved from frequency domain analysis compared to that obtained from corresponding time domain measurements. Development of Fourier transform (FT) Raman spectroscopy [1,2], Fourier Transform Infrared Spectroscopy (FTIR) [3], Fourier Transform Nuclear Magnetic Resonance (FTNMR) [4] are examples which are still in the process of further refinement. In the field of ultrasonics such attempts are rare though present day research on acoustic characterization of materials demand better accuracy and precision. More specifically, accuracy in the measurement of ultrasonic attenuation constant ($\alpha$) is relatively poor and depends very much on experimental conditions. The dependence of $\alpha$ on experimental condition has been explained clearly in recent theoretical studies [5]. It is shown that measured attenuation $\alpha_m$ will be, in general, higher than the true or intrinsic attenuation inside the medium. In measuring $\alpha$ for liquids, Fourier transform method was first used by C. Barnes et.al. [6]. The method used Fourier analysis of the time-domain response of the system to a single acoustic pulse. B Zhao et. al. [7] used a short time Fourier transform (STFT) method to estimate $v$ and $\alpha$. Two signals, viz., two echoes in pulse-echo method or two pulses in thru-transmission method are Fourier transformed and the amplitude spectra of the two are used for $\alpha$ calculation and phase spectra are used for calculating $v$. Though this method has some advantages, the accuracy is not better than other traditional spectrum methods. Similar techniques were also used by several other working groups [8-12] for various ultrasonic applications. These techniques are primarily based on the comparison of amplitudes and phases of the Fourier components of two signals, either the input and its echo, or two successive echoes, calculated separately. Selection of smaller time domain just to include the signal or the echo itself causes loss of information in the spectral data regarding frequency and corresponding amplitude leading to erroneous result. Specific information of the frequency spectrum with minute details of the component frequencies can be obtained from the Fourier analysis of the complete echo-train captured over a larger time domain. A computational parameter fitting method on the theoretically calculated and experimentally determined Fourier amplitudes of the complete echo-train gives much more fruitful results [13], but, being a parameter fitting method, it has limitations regarding the choice of input values of the desired parameters and a wrong choice may lead to unrealistic output values. This possibility is ruled out if the parameters can be extracted from the Fourier spectrum of the echo-train itself. The present report describes such a method for easier and convenient determination of the parameters v and α. The experimental procedure is described in sec. 2, sec. 3 gives results and discussion and conclusion is given in sec. 4.

## 2. The Experiment:

The experiment is done at room temperature maintained at $25^0 (\pm 1^0)$ C. The liquid sample is taken in a cylindrical glass container. A perfectly plain steel(glass) plate is introduced at the bottom of the container which acts as a reflector for the ultrasound signal. The transducer is mounted keeping its face parallel to the reflector surface on one end of a horizontal rod. The other end of the rod is attached to a vernier scale such that the transducer can move vertically and the length of liquid column can be measured with an accuracy of 0.002 cm. Ultran HE900 High Energy RF burst generator is used to generate electrical signal which is fed to the transducer placed at the top of the container with its face dipped into the liquid sample. RF bursts of 1 MHz and 2 MHz centre frequency with repetition time ~10 ms and p-p voltage ~200 V is used as input to the transducer. YOKOGAWA DL 1640 200MS/s 200MHz digital oscilloscope is used to capture echo train in the average mode to reduce noise. Data acquisition time is kept large such that the complete echo signal starting from the largest one to the smallest visible echo is contained well within the time span. This is to ensure finer details of the frequency components in the Fourier spectrum. Depending on the length (l) of the sample through which the pulse signal propagates, amplitude peaks appear in the frequency domain at certain frequencies f. These peaks are perfectly equidistant with spacing

$$\Delta f = v/2\, l \qquad (2.1)$$

The Q-value of the resonance peak at frequency f is related to α by the equation

$$\alpha = \pi f/Q \qquad (2.2)$$

Under favorable experimental condition, the peak frequency components are equidistant with high degree of accuracy and parameters measured from the FFT spectrum are more consistent and accurate. The method is demonstrated with Elix ultrapure de-ionized water.

## 3. Results and Discussion:

Typical pulse-echo train captured with ultrapure de-ionized water at 1 MHz signal frequency and its Fourier spectrum is shown in fig. 1 (a) and (b) respectively. The length (l) of water column for this case is l = 4.670 cm. Figure 1(b) shows the central part of the transducer bandwidth where the amplitudes of Fourier components are relatively high. The central resonance peaks are perfectly equidistant in frequency scale and from the frequency spacing Δv of resonance peaks, v is calculated using relation (2.1) and from the Q values of resonance peaks α is calculated using relation (2.2). The width of resonance peaks in the frequency spectrum is a consequence of wave attenuation in the medium. For an ideal loss free medium one expects δ-functions at resonance frequencies. In real systems, loss of energy at the interface comes into play in addition to the real energy loss during propagation in the medium. So measured attenuation $\alpha_m$ will be larger than real or intrinsic attenuation constant α. However, this effect can be reduced in the FSPE method by choosing the reflector material such that the reflection constant r at the interface is close to 1 and, at the same time increasing l such that only a few echoes are just detectable.

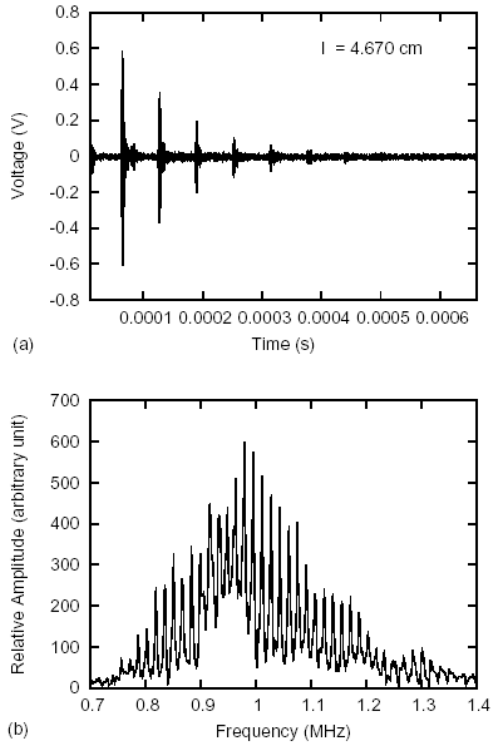

Figure 1: (a) Pulse echo signal with steel reflector for l = 4.670 cm and (b) its Fourier spectrum

The parameters v and $\alpha_m$ calculated for different l values at 1 MHz using steel reflector (r = 0.94) are given in Table-I. The results show that parameters calculated from FSPE obtained at different sample lengths agrees well with each other. In time domain pulse-echo measurement, pronounced dependence of $\alpha_m$ on l.

**Table-I:** FSPE values of parameters obtained for different l values at 1 MHz with steel reflector (r= 0.94).

| Length of water column (l) in cm | Velocity (v×10$^{-5}$) in cm/sec | Average velocity (v×10$^{-5}$) in cm/sec | Attenuation constant $\alpha_m$ in nepers/cm | Average attenuation constant $\alpha_m$ in nepers/cm |
|---|---|---|---|---|
| 4.862 | 1.484 | 1.486 ± 0.003 | 0.049 | 0.047 ± 0.008 |
| 5.140 | 1.491 | | 0.048 | |
| 5.514 | 1.487 | | 0.049 | |
| 5.704 | 1.484 | | 0.047 | |
| 5.994 | 1.485 | | 0.045 | |
| 6.016 | 1.486 | | 0.044 | |

has been reported earlier [13,14], but in the frequency domain FSPE measurement, the results agree quite well with each other. Values of the parameters obtained with glass reflector (r = 0.81) show similar results. Table-II summarizes values of v and $\alpha_m$ at 1 MHz and 2 MHz using steel and glass reflector.

**Table-II:** FSPE values of parameters at 1MHz and 2 MHz.

| Frequency in MHz | Steel reflector (r = 0.94) | | Glass reflector (r = 0.81) | |
| --- | --- | --- | --- | --- |
| | Velocity (v×10$^{-5}$) in cm/sec | Attenuation constant $\alpha_m$ in nepers/cm | Velocity (v×10$^{-5}$) in cm/sec | Attenuation constant $\alpha_m$ in nepers/cm |
| 1.0 | 1.486 ± 0.003 | 0.047 ± 0.008 | 1.502 ± 0.005 | 0.050 ± 0.009 |
| 2.0 | 1.484 ± 0.007 | 0.050 ± 0.008 | 1.508 ± 0.006 | 0.041 ± 0.007 |

### 4. Conclusion:

The prime concern of this report is frequency spectrum analysis of ultrasonic pulse-echo signal for the determination of ultrasonic propagation parameters v and α. It has been shown that the frequency spectrum of the pulse-echo signal captured over a longer time scale to include all of the echoes (starting from the largest one to the smallest detectable one) well within the time span of acquisition, show distinct and well-defined peaks corresponding to the resonance frequencies of the sample. Measurement of v and α from the separation and width of the resonance peaks are more consistent and accurate compared to the time domain measurement or other frequency domain measurements where Fourier components of isolated two signals, either the input and its echo, or two successive echoes, are considered. The attenuation value obtained in this method, viz., the frequency spectrum pulse-echo (FSPE) method, will be a little higher than the intrinsic attenuation value, because of the loss due to reflection at the boundary. But this error can be minimized using reflector material with high mechanical impedance to make r close to 1.

**Acknowledgement:** The author is grateful to Sankari Chakraborty and Papia Mondal for their assistance and technical help in conducting the experiment.